\documentclass[12pt,preprint]{aastex}

\usepackage{emulateapj5,apjfonts,endfloat} 

\begin{document}

\title{KECK SPECTROSCOPY OF 4 QSO HOST GALAXIES}

\author{J. S. Miller \altaffilmark{1} and A. I. Sheinis \altaffilmark{1}
}

\altaffiltext{1} {UCO/Lick Observatory, Astronomy Department, University
of California 1156 High Street, Santa Cruz, CA 95064}

\begin{abstract} We present optical spectroscopy  of the host galaxies
of 4 QSO's: PG 1444+407, PKS 2349-147, 3C 323.1, and 4C 31.63 having a
redshift range \( 0.1 \le z \le 0.3\). The spectra were obtained at the
Keck Observatory with the LRIS instrument offset 2-4 arcseconds from the
nucleus at several position angles in each galaxy. The objects close to
 3C 323.1 and PKS 2349-147 have the same redshifts of their
nearby QSOs and appear to be the nuclei of galaxies in the final states
of merging with the host galaxies. The spectra of the hosts show some
variety: PKS 2349-147 and 3C 323.1 show strong off-nuclear emission
lines plus stellar absorption features, while the other two show only
stellar absorption.  PKS 2349-147 and PG 1444+407 have a mixture of old
and moderately young stars, while 4C 31.63 has the spectrum of a normal
giant elliptical, which is very rare in our larger sample. The spectrum
of the host of 3C 323.1 appears to dominated by older stars, though our
data for it are of lower quality. The redshifts of the off-nucleus
emission lines and stellar components are very close to those of the
associated QSOs.

\end{abstract}


\section{INTRODUCTION } \label{s:intro}

A long history of imaging studies of QSO hosts dates back to early 60's
and 70's (\cite{MS66}, \cite{KRI73}). More recently there have been
several infrared (\cite{TDHR96}) and HST based imaging studies
(\cite{BKSS97}, \cite{MKDBOH99}). These studies have yielded colors,
morphologies and constraints on galaxy magnitudes. 

Spectroscopic observations are necessary for direct information on the
nature and kinematics of these nebulosities. Several groups have
performed spectroscopic investigations, dating back to the early 80's. 
The most extensive early investigations were those of
\cite{BO84} and \cite{BPO85}, who obtained off-nuclear spectra of 24
objects with the Palomar 200'' using a 2-arcsecond slit.  They
determined continuum colors for most objects and showed that a number of
them had an extended emission-line component, but, with the exception of
3C48, no stellar absorption lines were detected in any of the objects.
More recently several groups (\cite{H&H87}, \cite{HC90}, \cite{HKDB00}
and \cite{NDKHBJ00}) have made extensive observations with 4 meter class
telescopes.   With the advent of the Keck 10 meter telescope one group
has been able to compare detailed absorption spectra with population synthesis models for these objects
(\cite{CS00}, \cite{CS01}).

Conclusions about the nature of low-z QSO host galaxies differ. One
group of collaborators, (\cite{MKDBOH99}, \cite{HKDB00} and
\cite{NDKHBJ00}) believes these objects to be predominantly normal
massive ellipticals, while \cite{mil81}, in the first spectroscopic
investigation of a sample of these objects, concluded that they are not
normal luminous ellipticals, a result which was later confirmed with
better data from the Keck telescope in \cite{mts96}. Moreover, other
groups   ( \cite{CS00}, \cite{CS01} and \cite{she01}) have seen
post-starburst spectra in a substantial fraction of these objects.

In this Letter we present data from 4 of the 20 objects studied by us
with the Keck 10 meter telescope; a subsequent, more lengthy paper on
the entire sample is in preparation (\cite{S&M03}). The objects
discussed here- PG 1444+407, PKS 2349-147, 3C 323.1 and 4C 31.63-
exhibit a redshift of \( 0.1 \le z \le 0.3\).  Data acquisition and
reduction techniques are discussed in Section II, and Section III
presents a discussion of the results.

\section{DATA ACQUISITION AND REDUCTION} \label{s:aquisition and
reduction}

The data were taken using the Low Resolution Imaging Spectrograph (LRIS)
(\cite{OCCDHLLSS94}) on the Keck 10 meter telescope, and the observations
are summarized in table 1.  In general, a short observation of the QSO
(~$ 120$ sec) was followed by several longer exposures of the
off-nucleus regions, typically two to four exposures of 1800 sec at each
position. In most cases the slit location and orientation was determined
using published HST images \citep{BKSS97} to identify regions of high
surface brightness. These slit locations were mostly between 2 and 4
arcseconds  from the nucleus. The data were reduced with the VISTA data
reduction package \cite{stover88}, and final spectra were extracted by
standard techniques.  Typically 3 to 4 arcseconds along the slit
centered on the closest approach of the slit to the QSO were used for
the host, and since the slit was approximately 7 arcminutes projected on
the sky, it was straightforward to identify suitable regions along the
slit free of any host contamination for sky subtraction.  The last step
was correction for scattered QSO light, which is described in the next
section.


\begin{table} 
\begin{center} 
\caption{List of QSO Host Observations}

\vskip 8pt \begin{tabular}{llllclll} \hline\hline
\multicolumn{1}{c}{Object} & $z$ & date & Exposure & Slit width &
offset& P.A. & comment\\

\hline

PG1444+407& 0.267 & 8/96 & $2 \times 1800$& 1.0 & 3.0 & South @90& \\ & &
8/96& $ 2 \times 1800 $  & 1.0 & 2.8 & Northeast @135 &  \\ & & 7/97 & $ 3
\times 1800 $ & 1.0 & 3.0 & South@90 &\\

PKS2349-147& 0.173 & 8/96 & $2 \times 1800$ & 1.0 & 3.0 & South@90 &\\ & &8/96
& $ 2 \times 1800 $  & 1.0 & 2.0 & Southeast @20 &``LMC'' object \\ & &8/96 &
$ 2 \times 1800 $ & 1.0 & 4.0 & North@70 & North ``wisps''   \\

3C323.1 & 0.267 & 8/96 & $2 \times 1800$ & 1.0 & 2.0 & Southeast@35&\\ & &
8/96 & $ 2 \times 1800$ & 1.0 & 2.0 & Northwest@25& companion \\ & & 8/96 &
$1 \times 1800$ & 1.0 & 2.0 & Southeast@25&\\ & & 7/97 & $ 2 \times 1800 $ & 1.0
&2.8 & Southeast @35 &\\ & & 7/97 & $ 1 \times 1800 $ & 1.0 & 2.0 &
Southeast@35 &\\ & & 9/97 & $ 3 \times 1800 $ & 1.0 & 3.0 & North @90 & \\ & &
9/97 &$ 3 \times 1800 $ & 1.0 & 3.5 & South  @90 & \\

4C31.63 & 0.297 & 8/96 & $2 \times  1800$ & 1.0 & 4.5 & Northeast@135& \\ &&
8/96 & $ 4 \times 1800 $  & 1.0 & 3.0 & East @0 & \\ & & 7/97 & $ 3 \times
1800 $  & 1.0 & 2.0 & North @270 & \\ & & 7/97 & $ 2 \times 1800 $  & 1.0 & 2.5
&East @180 & \\ & & 9/97 & $ 2 \times 1800 $  & 1.0 & 3.0 & South @90 & \\ &
&9/97 & $ 3 \times 1800 $  & 1.0 & 3.0 & South @90 & \\

\hline\hline \end{tabular} \end{center}

\end{table}


\section{SCATTERED LIGHT SUBTRACTION}

The light from the QSOs in this study is 2-5 magnitudes brighter than
the entire host galaxy, which complicates the data reduction .  The
regions of the host galaxies observed in our study were typically 5 to 8
magnitudes fainter than the QSO. As a result of the blurring produced by
the earth's atmosphere plus a small contribution from the intrinsic
optical point-spread function and diffraction produced by the telescope,
some light from the QSO entered the spectrograph in the off-nucleus
observations.  To make an approximate correction for the scattered
light, the following process was adopted; it is very similar to that
used by \cite{BO84}. With the assumption that any broad Balmer lines
that are visible in the off-nucleus spectrum arise from scattered QSO
light, it is possible to estimate the fraction of scattered QSO light
present at the wavelengths of each of the broad lines.  The scattering
fractions are fit by a polynomial, and the spectrum of the QSO is scaled
by this scattering-efficiency polynomial and subtracted from the
off-nuclear spectra. The result is a spectrum with the broad lines and
other QSO spectral features removed.  The correction becomes uncertain
at wavelengths beyond the range covered by the Balmer lines. The
correction is especially uncertain  toward the blue end of the spectrum,
well beyond any detectable broad lines.

As a result of this uncertainty in the blue end of our spectra, we had
to adopt a more complex technique for the scatter subtraction for the
very bluest object, 3C 323.1.  In this approach we simultaneously
derive a model for the scattered QSO fraction and stellar and gas
components, assuming that the observed light can be represented entirely
by these three components. The model contains: (1) A two-age-component
stellar population synthesis model (\cite{BC1993}); (2) A model
spectrum for the narrow Balmer emission-line gas and (3) A scattered light model
which is the QSO spectrum multiplied by a
scatter efficiency curve. The scatter efficiency curve, $\xi$ has the
form: $ \xi = A_{1}  \lambda^{-\alpha} + A_{2} \lambda + A_{3}
\lambda^{2} $ where, $\alpha$ typically ranges from 1 to 10. All three model 
components are reddened equally. 

\section{DESCRIPTIONS OF INDIVIDUAL OBJECTS} \label{s:objects}

\subsection{PKS 2349-014}

\cite{BKS295} and \cite{BKSS97} present images of this object that show
obvious morphological signs of gravitational interaction, such as large
tidal arms and an extensive (50kpc) diffuse, off-center nebulosity. The
QSO has a typical spectrum with strong broad permitted lines and much
narrower forbidden lines with a modest amount of Fe II emission. They
identify several distinct regions for which we have obtained spectra.
Figure 1a was taken  3 arcseconds south of the nucleus in the region of 
the off-center diffuse nebulosity and illustrates 
the method of scattered light subtraction used for
all the spectra.  Spectrum A  includes the galactic light and the
scattered QSO light. Spectrum C is the QSO spectrum modeled and scaled
to represent the estimate of the scattered QSO component in Spectrum A,
and Spectrum B is the difference of A and C. In Spectrum B, the
absorption features show a strong resemblance to those of an elliptical
galaxy with superposed narrow emission lines. Figure 1b is the spectrum
of the so-called ``north wisps'', taken at 4 arcseconds north of the
nucleus. Comparison of these spectra with model spectra created from a
two-component population synthesis model (\cite{BC1993}) suggests
post-starburst spectra consisting of roughly 1/3 light from young stars
and 2/3 light from old (age $\geq$ 10 Gyr) stars. While these comparisons
with the population synthesis models show approximately the same
contribution to the light from young stars for each observation, they suggest a somewhat
younger population (age $\leq$ 500 Myr) for the observations to the
south of the host as compared to those of the ``north wisps'' (age
$\simeq$ 1 Gyr). Figure 1c shows the spectrum of the ``north wisps''
co-added with the other off-nuclear observation of the host galaxy from 
3 arcseconds south. These two positions comprise all of the off-nuclear
observations of this object that do not contain light from the companion
object. They have been co-added to increase the signal-to- noise. Figure
1d shows the companion located 1.8 arcseconds from the QSO.
\citep{BKSS97} note that this object has a luminosity similar to that of
the LMC if it is at the distance of the QSO and that it shows no sign of
interaction in the HST imaging. This spectrum shows an old stellar
population at the same redshift as the QSO, along with strong narrow
permitted and forbidden emission lines. Examination of the spectral data
as a function of position along the slit indicates that this emission is
not directly associated with the companion object, but is distributed
diffusely over the entire region. These data support the Bahcall et al.
speculation that this object is the tidally-stripped nucleus of a galaxy
about to be fully accreted by the host galaxy of PKS 2349-014.

\subsection{4C 31.63}

Fig 2a  shows the combined spectra for several observations to the east
of the nucleus of 4C 31.63, all of which appear identical to within the
accuracy of the data.  These observations are listed in table 1. This
radio-loud object has the spectrum of a typical  strong-Fe II QSO, with
strong Fe II emission along with the usual broad Balmer lines and very
weak or non-detectable forbidden lines. The off-nucleus spectrum shows
no detectable emission lines, but clearly shows the absorption features
of an old stellar population, i.e., the $4000 \AA$ break along with the
Ca II H and K, G band, and MgIb absorption features. The spectrum bears
a strong resemblance to that of a normal giant elliptical at the same
redshift as the QSO. Comparison of this spectrum to population synthesis 
models suggest all of the light is from an old stellar population (age $\geq$ 10 Gyr).

\subsection{PG 1444+407}

Fig 2b shows the combined spectra for all observations listed in table 1
of PG 1444+414 with the scattered light subtracted.  The spectrum of
this radio-quiet QSO is similar to that of 4C 31.63, showing strong Fe
II emission, broad Balmer lines and no detectable forbidden emission
lines. Again as with 4C 31.63, no emission lines are detected in the
off-nucleus spectrum.  In fact, in our entire sample of 20 QSOs, the
only ones that do not show off-nucleus emission are the strong Fe II
objects. Clearly seen are stellar absorption features, but the
absorption spectrum is different from that of the galaxy associated with
4C 31.63.  While some features produced by cool stars are again
apparent, such as  Ca II H and K and a weak G band, compared to the 4C
31.63 galaxy the $\lambda = 4000 \AA$ break is relatively weak, and the
spectrum longward of $4000 \AA$ is not as red; in addition some weak 
higher-order Balmer lines are visible. This is clearly not the
spectrum of a normal giant elliptical, as it looks like the combination
of a substantial fraction of relatively younger stars mixed in with the
older population. Comparison of this spectrum to population synthesis
models suggest this spectrum is composed of equal luminosity components of
very young stars (age $\leq$ 50 Myr) and old stars (age $\geq$ 10 Gyr), 
indicative of very recent, or ongoing star formation.

\subsection{3C 323.1}

The radio-loud QSO 3C 323.1 has a spectrum very typical of that of many
QSOs, with strong broad Balmer emission lines along with narrow
forbidden lines of [OII],[OIII], [NeIII], etc. There is a companion
object 2.1 arcsecond from the QSO, which is about 3.9 mag fainter than
the QSO itself \cite{BKSS97}.  Figure 2c shows a spectrum of this
companion. The companion has the same redshift as that of the QSO and a
spectrum with absorption features of cool stars as indicated plus [O
III] emission.  The [O III] emission is extended along the slit in the
original data and does not appear to be associated with the companion
itself.  Consistent with the conclusions of \cite{CS97}, the absorption
spectrum is not that of a normal giant elliptical, as the $4000 \AA$
break is not as pronounced and the continuum is less inflected.  The
data are consistent with the companion being the remnant nucleus of a
galaxy in the final stages of merging as suggested by \cite{stockton82}.

Figure 2d shows the spectrum of the nebulosity around  3C 323.1 with the
scattered light removed. This spectrum was formed by combining
observations taken at several different position angles as listed in
table 1, positions that avoided the companion. The data from the various
slit positions all appear very similar, and the combined spectrum was
formed to increase the signal-to-noise ratio. This spectrum shows very
weak $H_{\beta}$ in emission along with stronger narrow forbidden lines
of [OIII]. There appears to be a hint of the $4000 \AA$ break, but no
other absorption features are clearly visible. The hint of the $4000
\AA$ break and the overall spectral energy distribution suggests that
cool stars may be the main contributor to the observed light, but higher
quality data are needed for this faint galaxy component before anything
definitive can be said.

\section{DISCUSSION AND CONCLUSIONS}

\cite{BPO85} found correlations between the spectra of nebulosities
associated with QSOs and the spectra of the QSOs themselves, and similar
correlations are seen in our data.   Specifically, in our study the two
objects showing strong FeII emission, PG 1444+414 and 4C 31.63, show no
emission lines in the host galaxy  and have narrower, smoother, more
symmetric permitted lines in the QSO. One has a compact nuclear radio
source, while the other is radio quiet.  The other two objects, PG
2349-014 and 3C 323.1, show strong forbidden emission lines, broader
bumpier permitted emission lines, weak [Fe II] emission in the QSO,
and very extended emission regions in the hosts. These objects also
exhibit extended, steep radio structure.

The four QSOs we present in this study show a range in properties for
the host galaxies.  Only one host, that of 4C 31.63, has the spectrum of
a normal giant elliptical galaxy. Two objects have what appear to be the
remnant nucleus of a merging galaxy in the final stages of the merger
very close to nucleus of the host galaxy.  While close nuclear
companions are not that common, the other characteristics of the host
galaxies in these four objects are representative of our larger sample
of 20 QSOs.  Spectra typical of normal giant elliptical galaxies,
without strong extended emission line gas, are a distinct minority. Only
the spectrum of one additional object in our larger sample is consistent with this
type of galaxy.  Most of the hosts do show strong extended emission such
as that shown here in 3C 323.1 and PKS 2349-014. 

Data presented herein were obtained at the W.M. Keck Observatory, which is
operated as a scientific partnership among the California Institute of
Technology, the University of California and the National Aeronautics and
Space Administration. The Observatory was made possible by the generous
financial support of the W.M. Keck Foundation.

\acknowledgements

\bibliographystyle{apj} 




\begin{figure}
\plotone{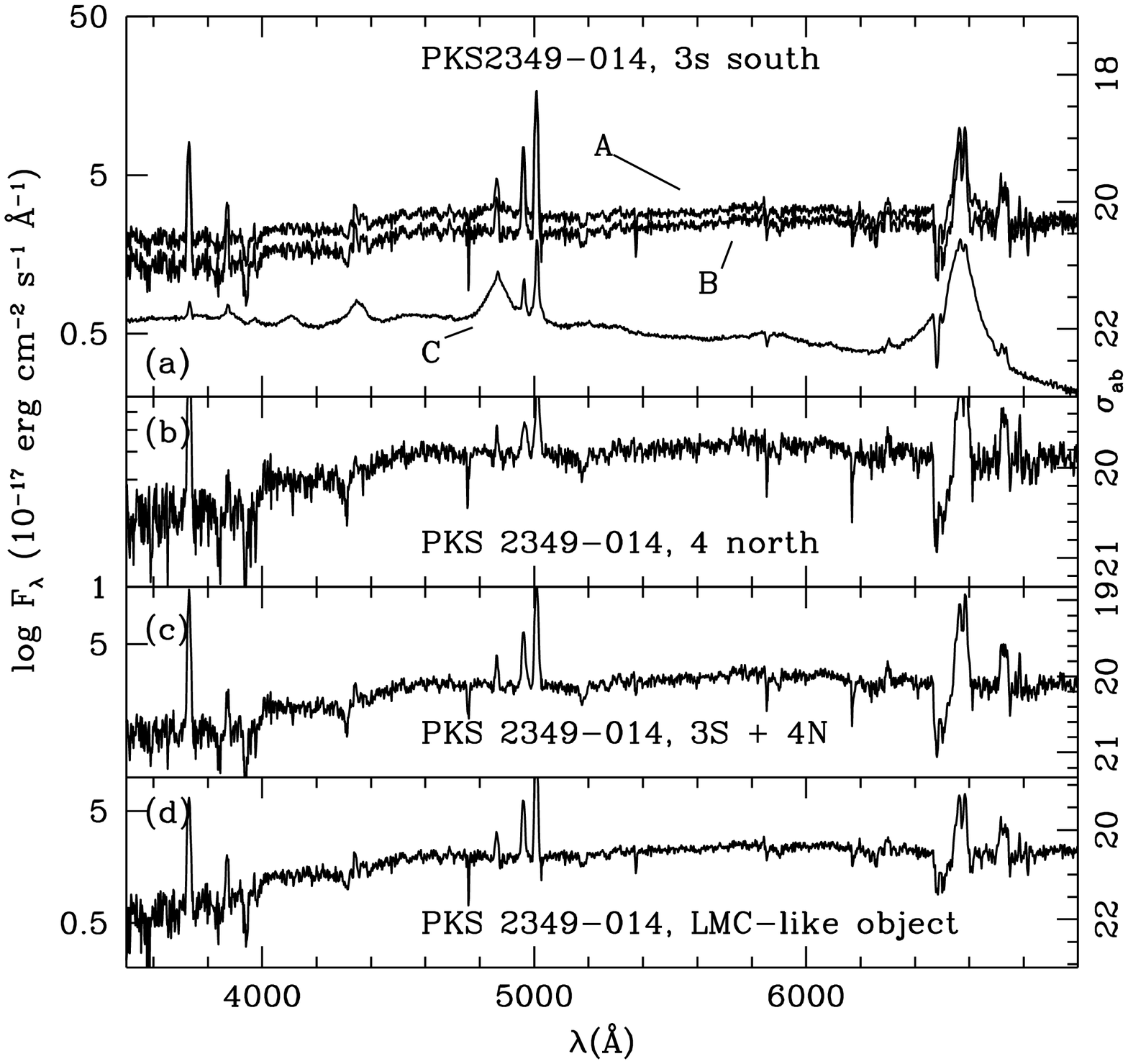}
\caption{Four spectra of PKS2349-014: top frame (a) shows the spectra of
PKS2349-014 taken at 3 arcseconds south of the nucleus: top curve (A) is
the extracted spectrum, middle curve (B) is the residual host galaxy
spectrum all position angles co-added, with scattered light subtracted
and the bottom curve (C) is the subtracted scatter model ; second frame
(b) is the residual host galaxy spectrum taken at 4 arcseconds north of the
nucleus; third frame (c) is the co-added spectra from 3 arcseconds south and
4 arcseconds north; bottom frame (d) is the spectrum from the ``LMC-like''
companion to PKS2349-014.  The increased noise at the red end is due to
strong OH night-sky lines.}
\end{figure}




\begin{figure}
\plotone{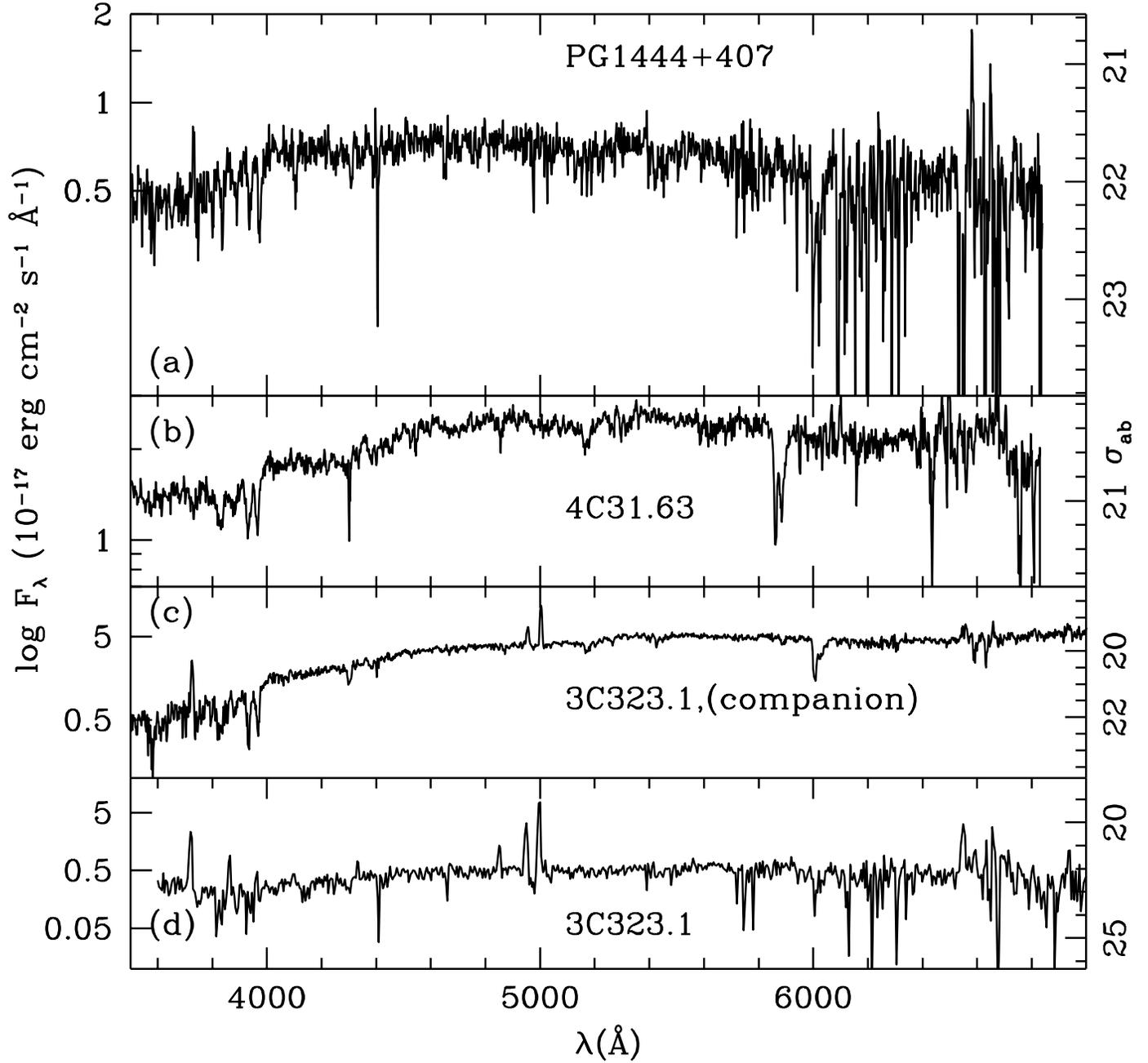}
\caption{Spectra of 3 objects:  top frame (a) shows the spectra of
PG1444+407; second frame (b) is the residual host galaxy spectrum of
4C31.63; third frame (c) is the residual host galaxy spectrum of the
companion to 3C323.1; bottom frame (d) is the residual host galaxy
spectrum of 3C323.1.  The increased noise at the red end is due to
strong OH night-sky lines.}
\end{figure}




\end{document}